\documentclass[aps,twocolumn,prl,showpacs,superscriptaddress,groupedaddress]{revtex4}

\usepackage{dcolumn}
\usepackage{bm}

\def\ltape{\hbox{\ $<$\hskip -8pt\raise -4pt\hbox{$\sim$}\ }}
\def\gtape{\hbox{\ $>$\hskip -8pt\raise -4pt\hbox{$\sim$}\ }}

\pdfoutput=1

   \ifx\pdfoutput\undefined
   \usepackage[dvips]{graphicx}
   \else
   \usepackage[pdftex]{graphicx}
   \fi
   \usepackage{epstopdf}
   \DeclareGraphicsExtensions{.pdf,.gif,.jpg,.eps,.png}
  
   \usepackage{amsmath}
  \usepackage{amssymb}

\begin{document}


\title{Scaling of magnetic reconnection in relativistic collisionless plasmas}

\author{Yi-Hsin~Liu}
\affiliation{NASA-Goddard Space Flight Center, Greenbelt, MD 20771}
\author{Fan~Guo}
\affiliation{Los Alamos National Laboratory, Los Alamos, NM 87545}
\author{William~Daughton}
\affiliation{Los Alamos National Laboratory, Los Alamos, NM 87545}
\author{Hui~Li}
\affiliation{Los Alamos National Laboratory, Los Alamos, NM 87545}
\author{Michael~Hesse}
\affiliation{NASA-Goddard Space Flight Center, Greenbelt, MD 20771}
\date{\today}

\date{\today}

\begin{abstract}
Using fully kinetic simulations, we study the scaling of the inflow speed of collisionless magnetic reconnection from the non-relativistic to ultra-relativistic limit. In the anti-parallel configuration, the inflow speed increases with the upstream magnetization parameter $\sigma$ and approaches the light speed when $\sigma > O(100)$, leading to an enhanced reconnection rate.  In all regimes, the divergence of pressure tensor is the dominant term responsible for breaking the {\it frozen-in} condition at the x-line. The observed scaling agrees well with a simple model that accounts for the Lorentz contraction of the plasma passing through the diffusion region. The results demonstrate that the aspect ratio of the diffusion region remains $\sim 0.1$ in both the non-relativistic and relativistic limits.

\end{abstract}


\maketitle

{\it Introduction--}
Magnetic reconnection is a process that changes the topology of magnetic fields and often leads to an explosive release of magnetic energy in nature. It is thought to play a key role in many energetic phenomena in space, laboratory and astrophysical plasmas \cite{ji2011}. In recent years, relativistic reconnection has attracted increased attention for its potential of dissipating the magnetic energy and producing high-energy cosmic rays and emissions in magnetically dominated astrophysical systems \cite{hoshino12a}, such as pulsar winds \cite{coroniti90a,arons12a,lyubarsky01a}, gamma-ray bursters \cite{thompson94a,zhangB11a,mckinney12a} and jets from active galactic nuclei \cite{beckwith08a,giannios10a,jaroschek04a}. However, many of the key properties of magnetic reconnection in the relativistic regime remain poorly understood. While early work found the rate of relativistic magnetic reconnection may increase compared to the nonrelativistic case due to the enhanced inflow arising from the Lorentz contraction of plasma passing through the diffusion region \cite{blackman94a,lyutikov03a}, it was later pointed out that the thermal pressure within a pressure-balanced Harris sheet will constrain the outflow to mildly relativistic conditions where the Lorentz contraction is negligible \cite{lyubarsky05a}, and a relativistic inflow is therefore impossible. Recently, the role of temperature anisotropy \cite{tenbarge10a}, inflow plasma pressure \cite{zenitani09a}, two-fluid \cite{zenitani09a} and inertia effects \cite{comisso14a} have been considered. All of these theories are generalizations of the steady-state Sweet-Parker \cite{sweet58a,parker57a} or Petschek-type \cite{petschek64a} models, which do not account for the mechanism that localizes the diffusion region and determines the reconnection rate in collisionless plasmas. Meanwhile, a range of reconnection rates are reported in computational works with different simulation models and normalization definitions \cite{zenitani09a,bessho12a,cerutti12a,sironi14a,fan14a,melzani14a}. However, the scaling of the rate has yet to be determined and the kinetic physics of the diffusion region is poorly understood in the relativistic limit.

In this work, a series of two-dimensional (2D) full particle-in-cell (PIC) simulations have been performed to understand the properties of reconnection in the magnetically dominated regime. We analyze in detail the current sheet structure and determine the scaling of the reconnection inflow and normalized rate as functions of upstream magnetization parameter $\sigma$ and magnetic shear. For electron-positron pairs with mass ratio $m_i/m_e=1$, we define the magnetization parameter as the ratio of the magnetic energy density to the plasma energy density, $\sigma \equiv B^2/(8\pi w)$ with enthalpy $w=n'_e m_ec^2+[\Gamma/(\Gamma-1)]P'_e$. Here $\Gamma$ is the ratio of specific heats and $P'_e\equiv n'_e T'_e$ the plasma thermal pressure in the rest frame. The shear Alfv\'en speed is $V_A=c[\sigma/(1+\sigma)]^{1/2}$ \cite{zenitani09a, sakai80a,anile89a,lichnerowicz67a}. In this Letter, the primed quantities are measured in the fluid rest (proper) frame, while the unprimed quantities are measured in the simulation frame unless otherwise specified. 
As pointed out in Ref.~\cite{lyubarsky05a}, if a simple pressure balance $P' \sim B^2/8\pi$ is satisfied across the current sheet, this will constrain the effective $\sigma \sim O(1)$, and thus restrict the inflow speed $V_{in} \ll c$. While this argument applies to the initial evolution of Harris-type layers, at later times the structure upstream of the x-line becomes nearly force-free as the pressure within the initial layer is depleted. As a consequence, the effective $\sigma$ increases and relativistic inflows develop. This conclusion is supported by additional simulations that start with a force-free current sheet and quickly develop relativistic inflows.
Furthermore, we present a simple model that predicts the inflow speed and normalized reconnection rate in collisionless plasmas, and explains the observed critical value of the upstream magnetization parameter $\sigma > O(100)$ necessary for generating a relativistic inflow. On the other hand, it is well known that the normalized collisionless reconnection rate, $R=V_{in}/V_{Ax}\sim 0.1$, in the non-relativistic limit can be estimated by the aspect ratio of the diffusion region, but the precise physics that determines this value remains mysterious \cite{yhliu14a}. Here $V_{Ax}$ is the Alfv\'en wave velocity in the outflow direction. Interestingly, the simulation results in this study suggest that this aspect ratio of $\sim 0.1$ persists in the relativistic regime. In addition, the analysis of the generalized Ohm's law shows that the divergence of pressure tensor $\nabla\cdot \tensor{P}_e$ is the dominant term responsible for breaking the {\it frozen-in} condition during intervals when the x-line is quasi-steady. However, the diffusion region is often highly dynamic due to the repeated formation of secondary tearing modes similar to the non-relativistic limit \cite{daughton07a}. As these islands are generated, the time-derivative of the inertial term also contributes significantly to breaking the {\it frozen-in} condition.

{\it Simulation setup--}
Most cases discussed here start with a relativistic Harris sheet \cite{kirk03a,zenitani07a,wliu11a,bessho12a,melzani14a}. The initial magnetic field ${\bf B}=B_0 \mbox{tanh}(z/\lambda) \hat{\bf x}+ B_g \hat{\bf y}$ corresponds to a layer of half-thickness $\lambda$ with a shear angle $\phi=2\mbox{tan}^{-1}(B_0/B_g)$.
Each species has a distribution $f_h \propto \mbox{sech}^2(z/\lambda)\mbox{exp}[-\gamma_d(\gamma_Lmc^2+ mV_d u_y)/T']$ in the simulation frame, which is a component with a peak density $n'_0$ and temperature $T'$ boosted by a drift velocity $\pm V_d$ in the y-direction for positrons and electrons, respectively. Here ${\bf u}=\gamma_L {\bf v}$ is the 4-velocity, $\gamma_L=1/[1-(v/c)^2]^{1/2}$ is the Lorentz factor of a particle, and $\gamma_d \equiv 1/[1-(V_d/c)^2]^{1/2}$. The drift velocity is determined by Amp\'ere's law $cB_0/(4\pi\lambda)=2 e\gamma_d n'_0 V_d $.  
The temperature is determined by the pressure balance $B_0^2/(8\pi)=2 n'_0 T'$.
The resulting density in the simulation frame is $n_0=\gamma_d n'_0$.  In addition, a  non-drifting background component $f_b \propto \mbox{exp}(-\gamma_L m c^2/T_b)$ with a uniform density $n_b$ is included. 
The simulations are performed using VPIC \cite{bowers09a}, which solves the fully relativistic dynamics of particles and electromagnetic fields. 
Densities are normalized by the initial background density $n_b$, time is normalized by the plasma frequency $\omega_{pe}\equiv(4\pi n_b e^2/m_e)^{1/2}$, velocities are normalized by the light speed $c$, and spatial scales are normalized by the inertia length $d_e\equiv c/\omega_{pe}$. 
All simulations used more than 100 particles per cell for each species.
The boundary conditions are periodic in the x-direction, while in the z-direction the field boundary condition is conducting and the particles are reflected at the boundaries.  
A localized perturbation with amplitude $B_z=0.03B_0$ is used to induce a dominant x-line near the center of simulation domain. The simulation parameters for the various runs considered in this Letter are summarized in Table 1. Our primary focus in the following section is the case ``Harris-4'' which illustrates the dynamics in the transition to the limit with relativistic inflows ({\it i.e.,} $V_{in} \approx c$). The domain size is $L_x\times L_z=384d_e \times 384d_e$ with $3072\times6144$ cells. The half-thickness of the initial sheet is $\lambda=d_e$, $n_b=n'_0$, $T_b/m_ec^2=0.5$, $B_g=0$ and $\omega_{pe}/\Omega_{ce}=0.05$ where $\Omega_{ce}\equiv eB_0/(m_e c)$ is a cyclotron frequency. Since the upstream magnetization parameter is more relevant in the nonlinear phase, it is used to parametrize our runs. The contribution from the reconnecting component is $\sigma_x \equiv B_0^2/(8\pi w)=(\Omega_{ce}/\omega_{pe})^2/\{2[1+(\Gamma/\Gamma-1)(T_b/m c^2)]\}$, which is $88.9$ with $\Gamma=5/3$. For cases with $T_b/m_ec^2 > 1$ in Table 1, we will use $\Gamma=4/3$ \cite{weinberg72a,synge57a}. 

\begin{figure}
\includegraphics[width=8.5cm]{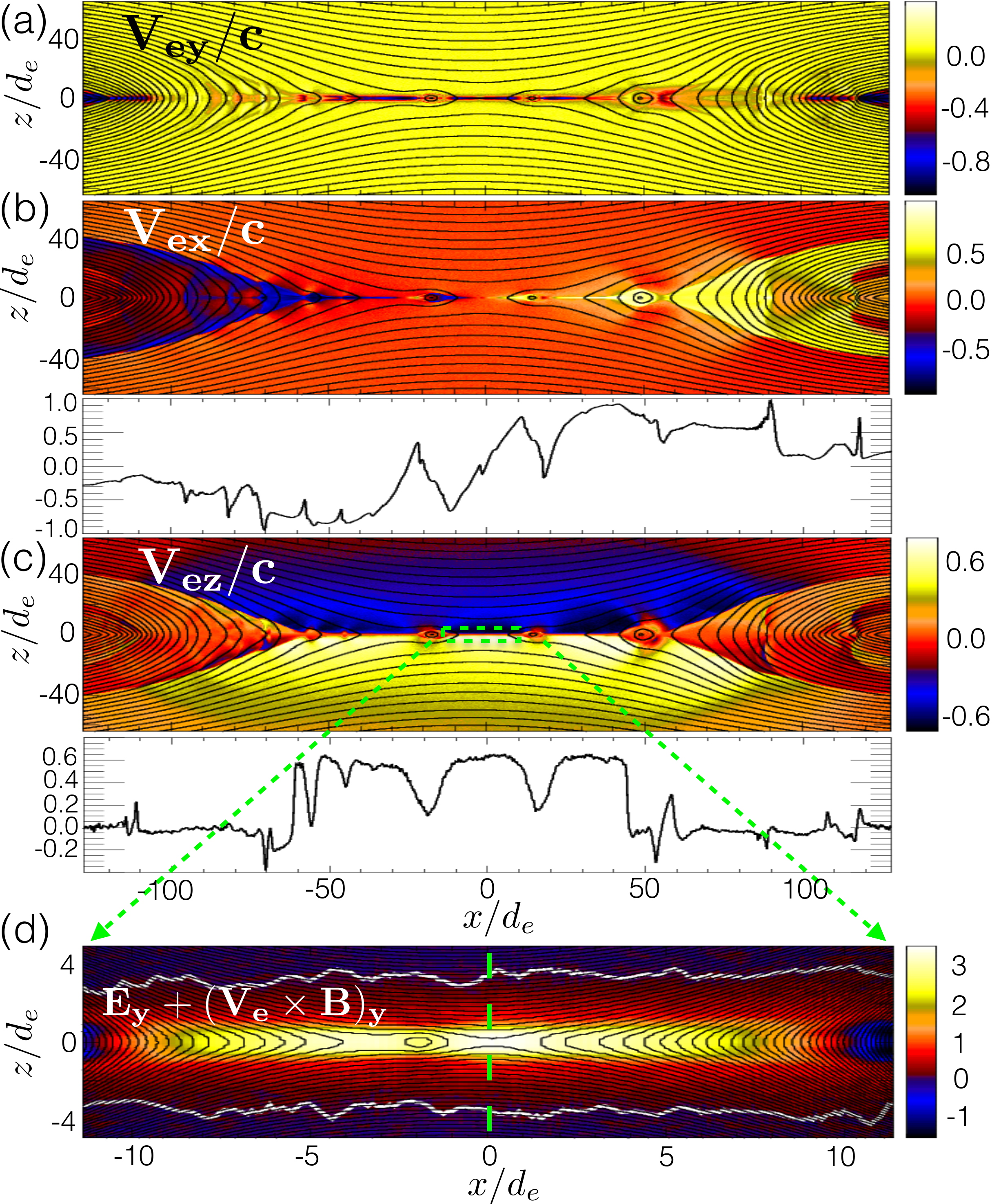} 
\caption {Results from the case ``Harris-4''  during the fully nonlinear phase at $566.4/\omega_{pe}$ showing (a) the electron out-of-plane speed $V_{ey}$, (b) outflow speed $V_{ex}$ with cut at $z=0$, (c) inflow speed $V_{ez}$ with cut at $z=-3.5d_e$, (d) a blow-up of the non-ideal electric field $E_y+({\bf V}_e\times {\bf B})_y$ inside the green-dashed box depicted in (c). The non-ideal electric field is positive in between the horizontal white curves. Black contours are flux surfaces in (a)-(d).} 
\end{figure}

{\it Simulation results--}
Fig.~1(a) shows the structure of current sheet in the nonlinear stage, where the current density concentrates within a layer with a half-thickness $\sim d_e$. This thickness appears to be independent of the initial sheet thickness, and scales with the inertia length based on the asymptotic background density ($n_b$). As shown in Fig.~1(b), the outflow velocity approaches the speed of light, while in Fig.~1(c) the peak inflow speed is $\sim 0.65c$. Note that these relativistic inflows also penetrate deeply across the magnetic separatrix into the $d_e$-scale sheet in the downstream region ($|x| \sim 50 d_e $). In addition, the simulation shows a rapid growth of secondary tearing modes, not only around the major x-line, but also along the concentrated current sheet that extends into the outflow exhausts. Interestingly, the secondary tearing mode appears considerably shorter in comparison with those in the non-relativistic regime. As shown in the blow-up (Fig. 1(d)), a magnetic island at $(x\sim-2d_e, z=0) $ is immersed inside the region where the {\it frozen-in} condition is broken, and it has a size $\sim 3 d_e \times 2 d_e$, implying that the secondary tearing mode grows for wave vectors $k_x \bar{\delta} > 1$. Here $\bar{\delta}$ is the half-thickness of the intense nonlinear current layer. In contrast, the initial tearing mode based on the relativistic Harris equilibrium is still constrained by $k_x \lambda < 1$ ({\it i.e.,} from relativistic energy principle) \cite{otto84a}, as in the non-relativistic limit. A temperature anisotropy developed in the nonlinear phase may change the stability criterion \cite{karimabadi04a}, however, to revolve this issue in the relativistic regime is beyond the scope of this Letter. Fig.~1(d) shows that the non-ideal electric field is also concentrated in a region $|z|< d_e$. However, the {\it frozen-in} condition starts to fail inside a wider layer in between the horizontal white curves, which may be due to a larger effective inertial scale based on a smaller density at $|z| \gtrsim d_e$ (see the density cut in Fig.~2(a)). Fig. 2(a) shows that the inflow velocity $V_{ez}$ reaches its maximum $\sim 0.65c$ at the location where {\it frozen-in} starts to fail ({\it i.e.,} marked by the green circle on the $E_y+({\bf V}_e \times {\bf B})_y$ curve). The profile of $V_{ez}$ is rather flat in between this location and $z=d_e$. Motivated by this observation, we use the local magnetic field $B_{x,u}$ at this location ($z\sim 3.5 d_e$) to normalize the reconnection electric field $E_y$, and the normalized electric field traces the evolution of the peak inflow velocity well (Fig.~2(b)), as expected.  Since in this relativistic regime $V_{Ax} \approx c$, these two quantities are both equivalent measurements of the normalized reconnection rate as discussed in the following section. The original peak density at the center of the sheet in the simulation frame is $n_0+n_b=\gamma_d+1\approx 11$. This peak density drops significantly from $11$ to $\sim 2$ and the density along the symmetry line ($z=0$) remains $\sim 2-4$, except inside secondary islands. The density ratio between the region immediate upstream to the x-line is $\sim2.5/0.3=8.3$. These numbers will be used to estimate the compression factor in the following section.  Per Ampe\'re's law, the density changes inside this $d_e$-scale layer require a reduction of the local magnetic field since the motion of the current carrier is limited by the light speed \cite{zenitani08b}. The double-peak in the density profile \cite{melzani14a} presumably comes from the Speiser orbits \cite{speiser65a}.  

To examine the mechanism of flux breaking, we employ the relativistic generalization of Ohm's law ${\bf E}+{\bf V}_e \times {\bf B}+(1/en_e)\nabla\cdot \tensor{P_e}+(m_e/e)(\partial_t {\bf U}_e+{\bf V}_e\cdot\nabla {\bf U}_e)=0$. Here ${\bf U} \equiv (1/n)\int d^3u {\bf u} f$ is the moment of 4-velocity. In contrast, the fluid velocity is ${\bf V} \equiv (1/n)\int d^3u{\bf v} f$. The pressure tensor, $\tensor{P} \equiv \int d^3u {\bf v u}f-n{\bf V}{\bf U}$, defined in this manner is consistent with that in the non-relativistic regime \cite{hesse07a,melzani14a}. 
Each term along the vertical cut in Fig.~1(d) is plotted in Fig.~2(c). There are strong oscillations in both $\nabla\cdot \tensor{P}_e$ and ${\bf V}_e\cdot\nabla{\bf U}_e$, which largely cancel each other. In comparison, the magnitude of the non-ideal electric field ${\bf E}_y+({\bf V}_e\times{\bf B})_y$ is much smaller. In Fig.~2(d), we examine the region around the neutral point, which demonstrates that ${\bf V}_e\cdot\nabla{\bf U}_e$ vanishes at $z=0$ since the neutral point coincides with the stagnation point in this symmetric configuration. The thermal pressure term, $\nabla\cdot \tensor{P}_e$, balances the non-ideal electric field at the x-line while the time-derivative  of the inertia $\partial_t {\bf U}_e$ remains small at this time \cite{hesse07a}. This is consistent with the study in the non-relativistic limit \cite{hesse99, hesse04,ricci04a,pritchett04}. However, the intense current layer is strongly unstable to secondary tearing modes. The time-derivative of inertia $\partial_t {\bf U}_e$ becomes finite positive when the $d_e$-scale current layer extends in length, and $\partial_t {\bf U}_e$ becomes finite negative (i.e., contributing to reconnection) when a secondary tearing starts to emerge in a sufficiently long layer.

\begin{figure}
\includegraphics[width=8.5cm]{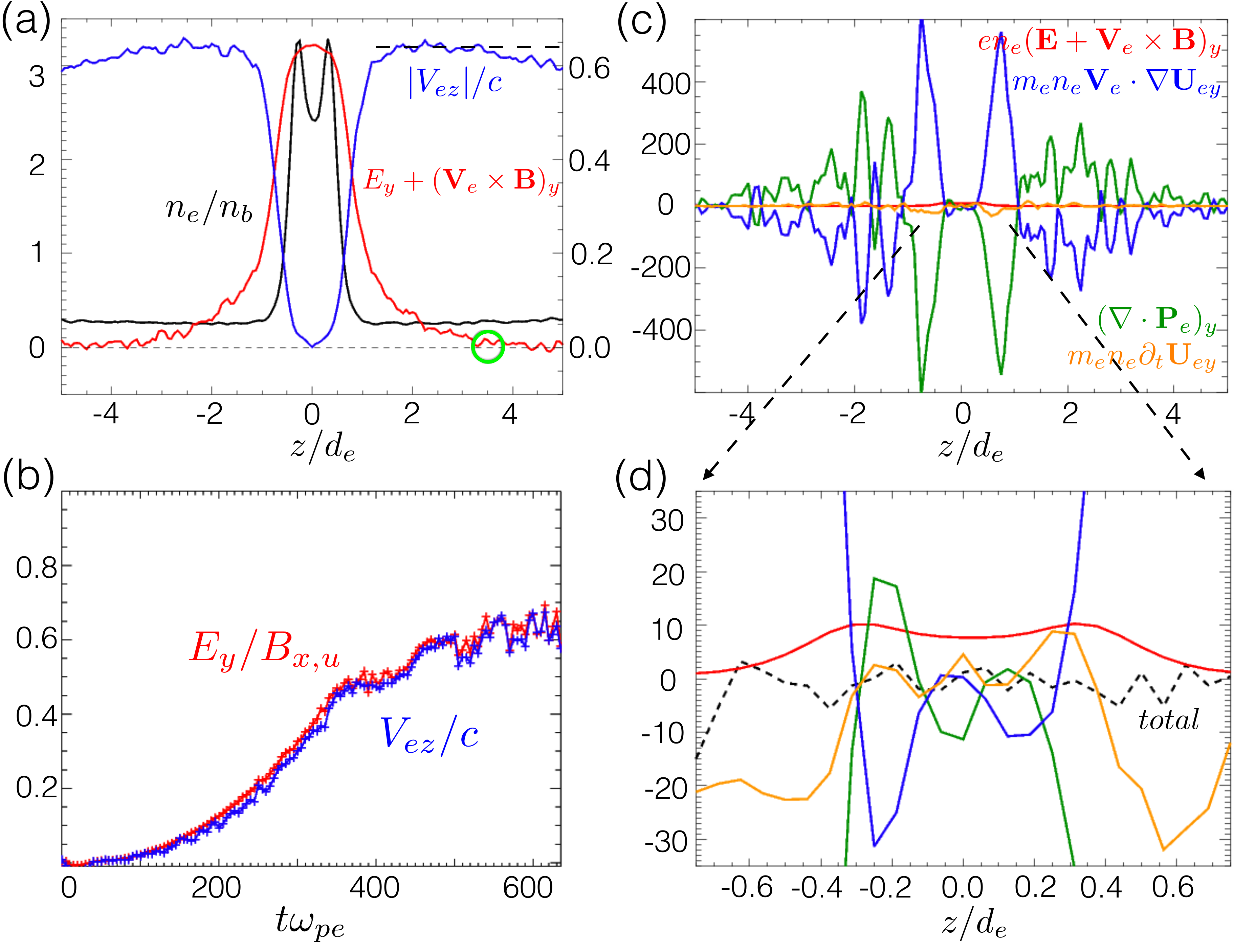} 
\caption {In (a), $n_e$, $E_y+({\bf V}_e\times {\bf B})_y$ and $|V_{ez}|$ along the vertical cut shown in Fig.~1(d). $|V_{ez}|$ is scaled by the right axis. The green circle marks the location where the {\it frozen-in} condition starts to fail; In (b), the evolution of the normalized reconnection electric field $E_y/B_{x,u}$ and the peak $V_{ez}$ near the major x-line at $\mbox{min}(A_y)$ along $z=0$. Here $A_y$ is the y-component of vector potential; In (c), quantities of Ohm's law along the vertical cut shown in  Fig.~1(d); In (d), the blow-up of (c) near the magnetic neutral point.} 
\end{figure}

{\it Simple theory--}
While previous theories \cite{blackman94a,lyubarsky05a,lyutikov03a,comisso14a} generalize the Sweet-Parker \cite{sweet58a,parker57a} or Petschek's \cite{petschek64a} models into relativistic regime, here we simply analyze the conservation of mass including the influence of the Lorentz contraction over a control-volume of size $L \times \delta$,

\begin{equation}
V_{in}\gamma_{in}n'_{in}L=V_{out}\gamma_{out}n'_{out}\delta.
\end{equation} 
Here the subscript ``{\it in}'' and ``{\it out}'' indicate the inflowing and outflowing plasmas, respectively. 
Given a magnetic shear angle, the outflow is limited by the upstream Alfve\'n wave velocity projected into the x-direction \cite{yhliu14a,melzani14a},
\begin{equation}
V_{out}=V_{Ax}=c\sqrt{\frac{\sigma_x}{1+\sigma}}.
\label{VAx}
\end{equation} 
Here the upstream magnetization parameter is $\sigma=\sigma_x+\sigma_g$ with $\sigma_g \equiv B_g^2/(8\pi w)$ accounting for the contribution from the guide field. The effective Lorentz factor based on the bulk flows is $\gamma_{out}=1/[1-(V_{out}/c)^2]^{1/2}=[(1+\sigma)/(1+\sigma_g)]^{1/2}$ and $\gamma_{in} =\{(1+\sigma)/[1+\sigma-\sigma_x(V_{in}/V_{Ax})^2]\}^{1/2}$. 

Working through the algebra, the peak inflow velocity can be determined with only one free parameter, $r_{n'}\delta/L$, 
\begin{equation}
\frac{V_{in}}{c}=\left(r_{n'}\frac{\delta}{L}\right)\sqrt{\frac{\sigma_x}{1+\sigma_g+(r_{n'}\delta/L)^2\sigma_x}},
\label{Vin}
\end{equation} 
where $r_{n'}\equiv n'_{out}/n'_{in}$ is the proper density ratio of the outflow to inflow. The compression factor is
\begin{equation}
\frac{n_{out}}{n_{in}}=r_{n'}\sqrt{\frac{1+\sigma}{1+\sigma_g+(r_{n'}\delta/L)^2\sigma_x}}.
\label{compression}
\end{equation} 
and the normalized reconnection rate is 
\begin{equation}
R\equiv \frac{V_{in}}{V_{Ax}}=\left(r_{n'}\frac{\delta}{L}\right)\sqrt{\frac{1+\sigma}{1+\sigma_g+(r_{n'}\delta/L)^2\sigma_x}}.
\label{R}
\end{equation} 
Note that $R$ differs from $V_{in}/c$ by a factor $c/V_{Ax}$.
From the {\it frozen-in} condition,  the normalized rate can also be written as $R=(c/V_{Ax})E_y/B_{x,u}$. In the limit of $V_{Ax} \rightarrow c$, then $R\sim V_{in}/c \sim E_y/B_{x,u}$ as shown in Fig.~2(b).

\begin{figure}
\includegraphics[width=8.5cm]{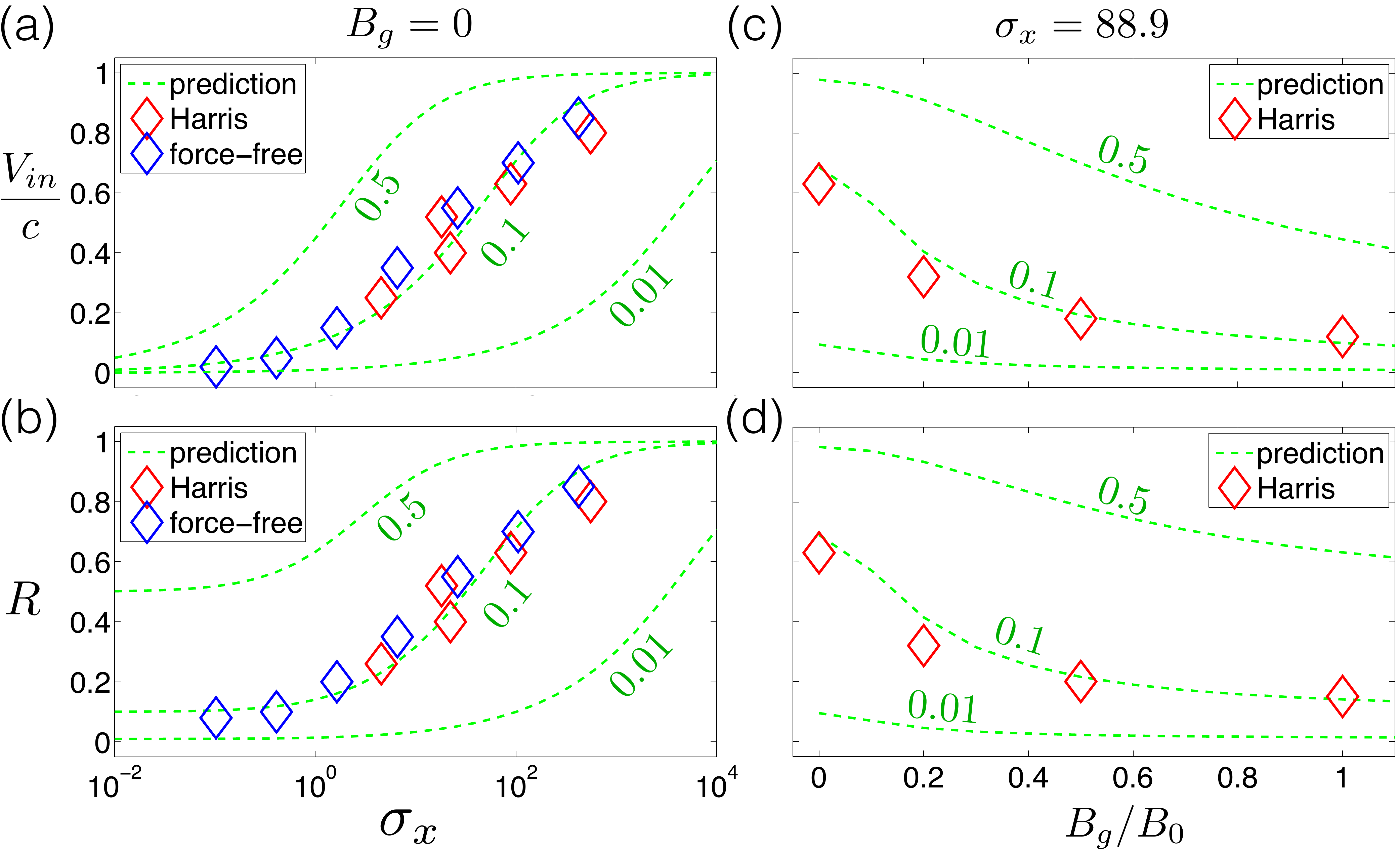} 
\caption {Scaling of the inflow velocity $V_{in}/c$  and the normalized reconnection rate $R$ as a function of $\sigma_x$ for cases with $B_g=0$ in the left; as a function of $B_g/B_0$ for cases with $\sigma_x=88.9$ in the right. Diamonds are measurements of runs in Table 1, green-dashed curves are predictions based on different value of $r_{n'}\delta/L$ as marked on the plots.} 
\end{figure}

With the assumption of $r_{n'}\delta/L=0.1$, as in the non-relativistic limit, Eqs. (\ref{Vin}) and (\ref{R}) immediately give $R \sim V_{in}/c =0.69$, consistent with the observed values for the case discussed. By comparing the measured compression factor $\sim 8.3$ in Fig.~2(a) and $n_{out}/n_{in}=6.9 r_{n'}$ from Eq (\ref{compression}), this implies that $r_{n'}\sim O(1)$ and therefore the aspect ratio $\delta/L\sim O(0.1)$. The aspect ratio of the intense $E_y+({\bf V}_e \times {\bf B})_y$ layer shown in Fig. ~2(d) seems to be consistent with this idea. To further test these predictions, a series of runs were performed over a wide range of parameters (listed in Table 1). The measurement of $V_{in}/c$ and $R$ are shown in Fig.~3 as diamonds, which agree closely with the predicted scaling based on $r_{n'}\delta/L=0.1$. This suggests that the aspect ratio of diffusion region persists during the transition from the non-relativistic to strongly relativistic regime.
With a larger $\sigma_x$, both the outflow and inflow speeds become closer to the light speed. For anti-parallel initial conditions ({\it i.e.,} $\sigma_g=0$), both $V_{in}/c$ and $R$ approach unity only when $\sigma_x > O(100)$, as shown in Fig.~3 (a)-(b), a condition obtained by demanding $(r_{n'}\delta/L)^2\sigma_x=0.01\sigma_x \gg1$ in the denominator of Eq. (\ref{Vin}) and (\ref{R}). On the other hand, with a guide field $B_g/B_0 \gtrsim O(1)$, the outflow speed (Eq.(\ref{VAx})) is slowed considerably below the light speed, the Lorentz contraction becomes negligible and the reconnection rate therefore goes back to $\sim 0.1$ as shown in Fig.~3(d). Our model appears to explain the scaling of the normalized rate observed in two-fluid simulations of Zenitani et. al. \cite{zenitani09a} as well. Unfortunately, due to the complexity of evolution in the nonlinear phase, we are not able to predict the reconnection rate normalized by the far upstream reconnecting component, which approaches a maximum of $\sim 0.3$ for cases with $\sigma_x\sim O(500)$ in the present study.


{\it Discussion--}
During the initial evolution of Harris-type current sheets, the pressure balance argument proposed in Ref. \cite{lyubarsky05a} does indeed restrict the inflow to $V_{in} \ll c$. However, as the new low pressure plasma is convected into the layer, the structure upstream of the x-line 
(except inside the intense $d_e$-scale layer) 
approaches the force-free limit and the inflows then become relativistic. During this later nearly force-free ({\it i.e.,} ${\bf J} \times {\bf B} \sim 0$) phase, the magnetic pressure gradient $\partial_z(B^2/8\pi)$ is balanced with the magnetic tension $({\bf B}\cdot \nabla B_z)/4\pi$, while the thermal pressure gradient is small.
To further test this idea, we have also performed an additional series of simulations using force-free current sheets as described in Guo et al. \cite{fan14a}. Due to the absence of the hot and dense sheet component, these force-free initial layers quickly develop relativistic inflows, consistent with the above argument. The measured $V_{in}/c$ and $R$ shown as blue diamonds in Fig.~3(a)-(b) follow the same trend, which further demonstrates that the scaling in the nonlinear stage is determined solely by the upstream parameters. 

In summary, a simple theory based on the Lorentz contraction and the assumption of an universal aspect ratio ($\sim 0.1$) of the diffusion region provide an explanation of why the inflows in anti-parallel reconnection become relativistic when the upstream magnetization parameter $\sigma$ exceeds $O(100)$. These results may be important for understanding particle acceleration in high-$\sigma$ regimes \cite{zenitani01a, fan14a} and to understand the dissipation of strong magnetic field in high-energy astrophysical systems, such as the ``$\sigma$-problem'' in the Crab Nebula \cite{kirk03a}, and the destruction of strong magnetic field near magnetars \cite{thompson94a} and black holes \cite{beckwith08a}.

\begin{table}[ht]
\caption{Parameters of Runs} 
\centering 
\begin{tabular}{c c c c c c c c c} 
\hline\hline 
Harris & 1 & 2 & 3 & 4 & 5 & 6 & 7 & 8 \\ [0.5ex] 
\hline 
$B_g/B_0$ & 0 & 0 & 0 & 0 & 0  & 0.2 & 0.5 & 1 \\ 
$n_b/n'_0$ & 1 & 0.25 & 1 & 1 & 1 & 1 & 1 & 1  \\
$T_b/mc^2$ & 2.5 & 2.5 & 0.5 & 0.5 & 0.5 & 0.5 & 0.5 & 0.5  \\
$\omega_{pe}/\Omega_{ce}$ & 0.1 & 0.05 & 0.1 & 0.05 & 0.02 & 0.05 & 0.05 & 0.05 \\
$\sigma_{x}$ & 4.5 & 18.2 & 22.2 & 88.9 & 555.6 & 88.9 & 88.9 & 88.9 \\[1ex]  
\hline\hline 
Force-Free & 1 & 2 & 3 & 4 & 5 & 6 & 7 \\ [0.5ex]
\hline
$B_g/B_0$ & 0 & 0 & 0 & 0 & 0 & 0 & 0  \\ 
$T_b/mc^2$ & 0.35 & 0.36 & 0.36 & 0.36 & 0.36 & 0.36 & 0.36  \\
$\omega_{pe}/\Omega_{ce}$ & 1.6 & 0.8 & 0.4 & 0.2 & 0.1 & 0.05 & 0.025 \\
$\sigma_x$ & 0.1 & 0.4 & 1.6 & 6.6 & 26.3 & 105.3 & 421 \\[1ex]  
\hline
\end{tabular}
\label{table:nonlin} 
\end{table}

\acknowledgments
Y.-H. Liu thanks for helpful discussions with S. Zenitani, N. Bessho and J. Tenbarge. We are grateful for support from NASA through the NPP program and the Heliophysics Theory program. Simulations were performed at the National Center for Computational Sciences at ORNL and with LANL institutional computing. 


\newpage

\end{document}